# The Impact of Large Erosional Events and Transient Normal Stress Changes on the Seismicity of Faults

**L. Jeandet Ribes[1,2], N. Cubas[1], H. S. Bhat[3], and P. Steer[2]**

[1] *Institut des Sciences de la Terre Paris, ISTeP UMR 7193, Sorbonne Université, CNRS-INSU, 75005 Paris, France*

[2] *Univ Rennes, CNRS, Géosciences Rennes - UMR 6118, 35000 Rennes, France*

[3] *Laboratoire de Géologie, Ecole Normale Supérieure, CNRS-UMR 8538, PSL Research University, Paris, France*

Corresponding author: Louise Jeandet Ribes (louise.jeandet@sorbonne-universite.fr)

**Key Points:**

- We investigate seismicity response to an erosional event by modelling the effects of transient normal stress changes on a frictional fault

- Erosional events with a duration shorter than a seismic cycle can increase the seismicity rate and the proportion of small earthquakes

- Large erosional events have the potential to contribute significantly to the deformation of the first kilometers of the Earth's crust

**Abstract**

The long-term erosion of steep landscapes is punctuated by dramatic erosional events that can remove significant amount of sediments within a time-scale shorter than a seismic cycle. However, the role of such large erosional events on seismicity is poorly understood. We use QDYN, a quasi-dynamic numerical model of earthquake cycles to investigate the effect of a large erosional event on seismicity. The progressive evacuation of landslide sediments is modelled by a transient normal stress decrease. We show that erosional events with a shorter duration compared with the duration of a seismic cycle can significantly increase the seismicity rate, even for small stress changes. Moreover, large erosional events with a shorter period compared with the earthquake nucleation time-scale can change earthquake size distribution by triggering more small events. Those results suggest that large erosional events can significantly affect seismicity, illustrating in turn the short-term impact of surface processes on tectonics.

## 1 Introduction

Over geological time scales, mountain belts classically grow through thrusting and thickening of the Earth's crust under tectonic forces (e.g., Davis et al., 1983). This long-term building results from deformation by viscous, ductile and brittle processes and by frictional slip along major faults, leading to rock uplift over a succession of seismic cycles (King et al., 1988; Le Béon et al., 2014). Mass transfers at the Earth's surface due to erosional processes imply stress changes at depth. According to numerical modeling, these stress changes partly control the size and long-term deformation of mountain ranges (Dahlen & Barr, 1989; Thieulot et al., 2014; Whipple, 2009; Willett, 1999). At shorter time scales (< 1Myrs), erosion and sedimentation are also suspected to affect fault slip rate (Calais et al., 2010; Cattin & Avouac, 2000; Theunissen & Huismans, 2019; Vernant et al., 2013).

At a seismic cycle time-scale (1-1000 years), mountain building is punctuated by rare but catastrophic tectonic and erosional events affecting the long-term landscape evolution. Succession of earthquakes induce permanent deformation (Simpson, 2015) and large erosional events represent a major contributor to long-term erosion rates (Kirchner et al., 2001; Marc et al., 2019). However, the potential influence of such sudden erosional processes on seismicity is still poorly understood.

The seasonal variations of snowload, precipitation, or atmospheric pressure are known to modulate static stresses at an annual time-scale (e.g., Heki, 2003). Although the variation of stress induced by these surface processes is small compared to earthquake stress drop (e.g., Shaw, 2013) or tectonic loading (e.g., Townend & Zoback, 2004), they do modulate the background seismicity along most tectonically active settings (Bollinger et al., 2007; Christiansen et al., 2007; Gao et al., 2000; Heki, 2003). The periodicity of these variations is likely a major parameter (Ader et al., 2014).

In mountainous areas, hillslopes regularly experience catastrophic erosional events triggered by large earthquakes or rainfall events. These sudden events, associated with numerous landslides, mobilize a large volume (up to several km$^3$) of sediments (Keefer, 1994; Marc et al., 2016) that will ultimately be evacuated by rivers. Using an elastic half-space model, Steer et al., 2014 proposed that the erosion rates of active tectonic settings such as Taiwan should be high enough to induce static stress variations of 0.01 to 1 MPa within the interseismic phase in the first few kilometers of the crust. This variation is suggested to be large enough to affect regional seismicity. However, the seismicity response to sudden erosional events is expected to strongly depend on the timing of evacuation of landslide-driven sediments. This time-scale is particularly difficult to constrain since many factors are in play (Croissant et al., 2019). These include landslide connectivity to the drainage network (Li et al., 2016), river dynamics (Croissant et al., 2017; Yanites et al., 2010), and the grain size distribution of landslide sediments (Cowie et al., 2008; Egholm et al., 2013; Sklar & Dietrich, 2006). This complexity led to estimations of evacuation timescales ranging from centuries (Stolle et al., 2018; Yanites et al., 2010) to only years to decades for suspended load (Hovius et al., 2011) or coarse sediment (Croissant et al., 2017; Howarth et al., 2012). In any case, this evacuation timescale is roughly smaller or equal to the typical duration of a seismic cycle (Chen et al., 2007; Shimazaki & Nakata, 1980; Sieh et al., 1989).

A relationship between a catastrophic erosional event and regional seismicity has been suggested for the typhon Morakot which struck Taiwan in 2008 (Steer et al., 2020). This typhoon triggered ~10000 landslides and removed about 1.2 km$^3$ of sediments from the hillslopes (Marc et al., 2019). The authors reported an increase of both earthquake frequency and b-value (i.e., an increase in the proportion of small earthquakes compared to large ones) directly following the typhoon and lasting for at least 2.5 years.

We here investigate if a stress change induced by the removal, over a certain duration, of the sediments following a sudden large erosional event could modify the seismicity of the neighboring crustal faults as suggested in Taiwan. Since this requires consideration of the fault response to transient shear stress increase, or normal stress decrease (Steer et al., 2014), it is necessary to account for the time-dependency of fault friction. Simple static stress change calculations offer limited comprehension of the problem (Ader et al., 2014). Therefore, we here use a numerical model considering the general case of a single fault embedded in an elastic medium obeying a rate-and-state friction law (Dieterich, 1979; Rice, 1993; Ruina, 1983). The fault is subjected to a normal stress decrease applied over a certain erosional time and we explore the resulting seismicity rate and earthquake size distribution.

## 2 Methods

We use QDYN (Luo et al., 2017), a boundary element model that simulates fault slip under a quasi-dynamic approximation (i.e., quasistatic-elasticity with radiation damping). Its adaptive time-stepping enables to simulate earthquake cycles including seismic and aseismic slip. We considered a 1-D, mode II fault embedded in an elastic medium (Fig.1). We assume our study also applies to reverse faults, since they are well approximated by a mode II rupture at depth, where interactions of the seismic waves with the free surface can be ignored (Madariaga, 2003; Oglesby et al., 1998).

The friction acting on the fault interface obeys a rate-and-state friction law (Marone, 1998):

$$\tau = \sigma_n \left[ \mu_0 + a \log \left( \frac{V}{V_0} \right) + b \log \left( \frac{\theta V_0}{D_c} \right) \right] \tag{1}$$

where $\tau$ is the shear strength, $\sigma_n$ the applied normal stress, $\mu_0$ the friction coefficient corresponding to the reference slip rate $V_0$, $\theta$ a state variable, and $D_c$ a characteristic slip distance for state variable evolution. The $a$ and $b$ parameters describe the rate and state dependencies, respectively. The state variable $\theta$ varies according to slip (Rice & Ruina, 1983). Laboratory experiments (Hong & Marone, 2005; Kilgore et al., 2017; Linker & Dieterich, 1992; Shreedharan et al., 2019) and theoretical analysis (Molinari & Perfettini, 2017) have shown that normal stress variations can contribute to the frictional state. However, since normal stress does not vary with slip in our models due to the flat fault geometry, we use the simple ageing law (Rice & Ruina, 1983) :

$$\dot{\theta} = 1 - \frac{V\theta}{D_c} \tag{2}$$

The fault is infinite in fault-perpendicular direction and includes a seismogenic patch with rate-weakening (RW) properties (a-b < 0) surrounded by two rate-strengthening (RS) areas (a-b > 0) of the same size (Figs 1a,b). The length of the seismogenic patch is set to be ten times the nucleation size (Rubin et al., 2005), which leads to a fault length of 17 km. The fault is discretized into cells of about 0.5 m in size to ensure the resolution of the cohesive zone (Lapusta et al., 2009) (see supporting information).

Frictional parameters and boundary conditions are set to commonly used values (e.g., Noda and Lapusta, 2010, Ader et al., 2014). The value of *b* is 0.014 and *a* varies from 0.02 in the RS domain to 0.01 in the RW zone (a/b = 0.7). The steady-state frictional properties are constant along the fault ($\mu_0$ = 0.6 and $V_0 = 10^{-9}$ m.s$^{-1}$) and the medium has a shear modulus of G = 30 GPa. The fault is loaded at a velocity $V_{pl}$ of 3 cm/year and the applied normal stress is of 10 MPa, consistent with a depth of a few kilometers (Suppe, 2014). Quasi-dynamic simulations of a seismogenic patch with constant frictional properties produces one characteristic, repeating event (Rice, 1993). Since multiple fault models are still under progress (Romanet et al., 2018), we choose to simulate a spatio-temporal complexity by varying the critical distance $D_c$ (Aochi & Ide, 2004; Hillers et al., 2007; Ide & Aochi, 2005). To obtain various earthquake magnitudes with a single fault, we vary $D_c$ along strike from values of $2 \times 10^{-5}$ to $3.4 \times 10^{-4}$ m following a self-similar pattern in both RW and RS patches (Fig. 1c).

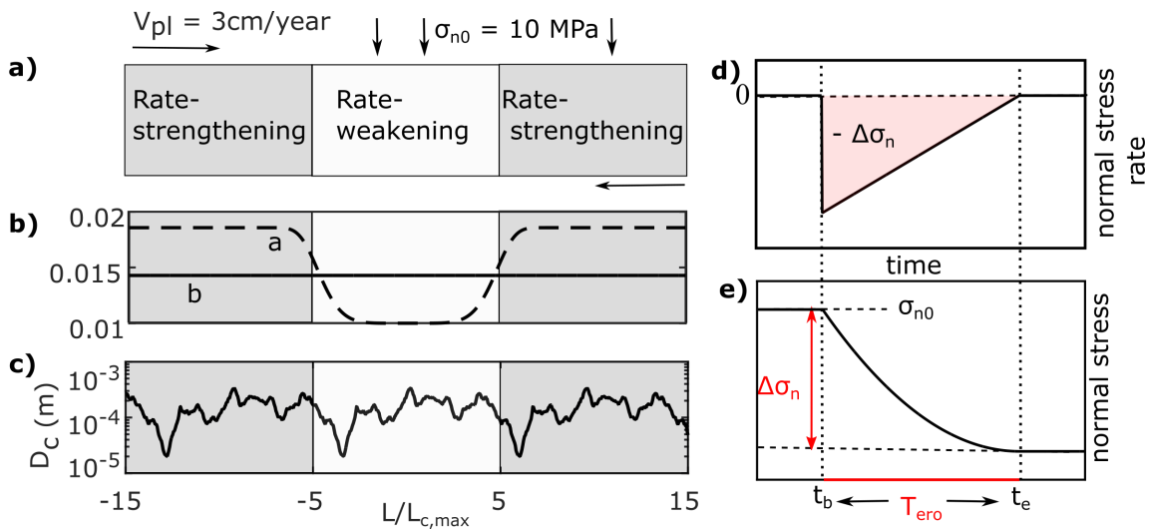

**Figure 1.** Numerical setup used in this study. a) Schematic of the simulated fault. Slip weakening acts over the central portion of the fault, of length $10 \times L_c$. The fault is loaded at a plate velocity $V_{pl}$ of 3cm/years and the normal stress $\sigma_n$ acts over the entire fault. b-c) Along-strike distribution of friction parameters (b) and critical distance $D_c$ (c). Normal stress rate (d) and normal stress (e) temporal variation implemented in QDYN to model one large erosional event. Before the erosional event, the normal stress is $\sigma_{n0}$. Erosion begins at $t_b$. A quantity $\Delta\sigma_n$ is removed over a period $T_{ero}$ until a new background value of normal stress is reached at $t_e$.

An erosional event is defined by the amplitude of the stress variation, its duration and the functional relationship of this variation. Inferred erosion-induced increase in coulomb stress ranges from 0.01 to 1 MPa (Steer et al., 2014) and estimates of the duration of an erosional event vary from 1-10 years (Croissant et al., 2017; Hovius et al., 2011), to several centuries (Stolle et al., 2018; Yanites et al., 2010). Moreover, a sharp erosion increase followed by a linear decrease down to its background value has been observed just after the Chi-Chi earthquake by Hovius et al. (2011).

We thus run simulations including a sudden drop in normal stress rate (Fig. 1d) followed by a linear increase taking place over a period $T_{ero}$, with a total removed normal stress integrated over $T_{ero}$ of $\Delta\sigma_n$ (Fig. 1e). We test $\Delta\sigma_n$ ranging from 0.01 to 1 MPa and $T_{ero}$ from $10^{-3}$ to 10 times the duration of one modelled seismic cycle (2.2 years). The corresponding mean normal stress rate thus varies from 6.34 Pa/s (for $\Delta\sigma_n = 1$ MPa, $T_{ero} = 0.01$ years) to $1.5 \times 10^{-5}$ Pa/s ($\Delta\sigma_n = 0.01$ MPa, $T_{ero} = 20.48$ years), i.e., between 5 and $10^{-5}$ times the background loading rate imposed by the plate velocity (~1.2 Pa/s). The onset of the normal stress perturbation is implemented during the interseismic period of a stabilized cycle (when the fault produces regular events). In the following, we use 'erosion' and 'normal stress decrease' to mean the same physical process.

For each simulation, we build an earthquake catalogue by isolating seismic events using a moment rate threshold $\dot{M}_0$ of $10^8$ dyn.cm$^{-2}$.s$^{-1}$ and we compute the magnitude of individual earthquakes assuming a fault width of 10 km (supporting information).

## 3 Results

Without any normal stress perturbation, we obtain a regular sequence composed of three characteristic earthquakes (Fig. 2a) that nucleate at an edge of the RW patch loaded by RS regions. The magnitude of these three typical events are of 4.85, 5.20 and 4.18, respectively (Fig. 3). The second event is a characteristic large earthquake that regularly ruptures the entire seismogenic area, with a recurrence time of 2.2 years. Therefore, in the following, the sequence is called "seismic cycle".

For a normal stress perturbation of $\Delta\sigma_n = 1$ MPa applied over a $T_{ero} = 0.08$ years, the seismicity rate increases during the erosional period (Fig. 2b). The new sequence starts with a large event ($M_w = 5.34$), that ruptures the entire patch. It is followed by a succession of earthquakes of various magnitudes (between 4.01 and 5.22), with some small events nucleating on the right portion of the fault, which is not characterized by small $D_c$ values (fig. 2d).

To characterize the size distribution of dynamic events, we arbitrarily bin the earthquakes generated during $T_{ero}$ into two categories: small ($M_w > 4.5$) and large $M_w < 4.5$ ruptures. We first compare two end-member simulations displaying different response in terms of earthquake magnitude (Fig. 3). For $T_{ero} = 20.48$ years, the earthquake frequency increases by a factor close to two during approximately ten years, and then progressively returns back to its initial level (Fig. 3a and c). The characteristic sequence is more frequent in time with the same magnitudes. For $T_{ero} = 0.01$ years, we observe a significant change in the distribution of earthquake magnitudes during erosion (Fig. 3d). Small events are increased by 60% and are more frequent than larger events (Fig. 3b).

We now show results obtained for simulations with $\Delta\sigma_n$ varying from 0.01 to 1 MPa and $T_{ero}$ from 7 hours to 20 years, corresponding to a ratio $T_{ero}/T_{cycle}$ ratio ranging from $4 \times 10^{-4}$ to 10 (Fig. 4). For each model, we plot the number of earthquakes N (Fig. 4a) and the cumulated moment (Fig. 4b) during the erosional period. We then compute the average earthquake frequency obtained during erosion (i.e., $N/T_{ero}$) normalized by the earthquake frequency observed during an undisturbed seismic cycle (i.e., $3/T_{cycle}$) (Fig. 4c). Earthquake statistics during the erosional event are also given, as a function of the ratio $T_{ero}$ over either the duration of a standard seismic cycle ($T_{cycle}$) of the undisturbed fault (Fig 4c) or the nucleation time ($T_{nuc}$, figure S4) of a characteristic earthquake (Fig 4d).

At first order, the number N of earthquakes during erosion increases with $\Delta\sigma_n$ (Fig. 4a). In turn, the cumulated seismic moment during the erosional period follows the same pattern (Fig. 4b). For $\Delta\sigma_n = 0.1$ and $0.01$ MPa, very low $T_{ero}$ are too short-lived to enable any triggering during the period of erosion. At second order, we can identify two different regimes depending on the duration of the erosional event. For $T_{ero}/T_{cycle}>1$ , N increases when increasing $T_{ero}$, whereas it remains roughly constant for $T_{ero}/T_{cycle}<1$.

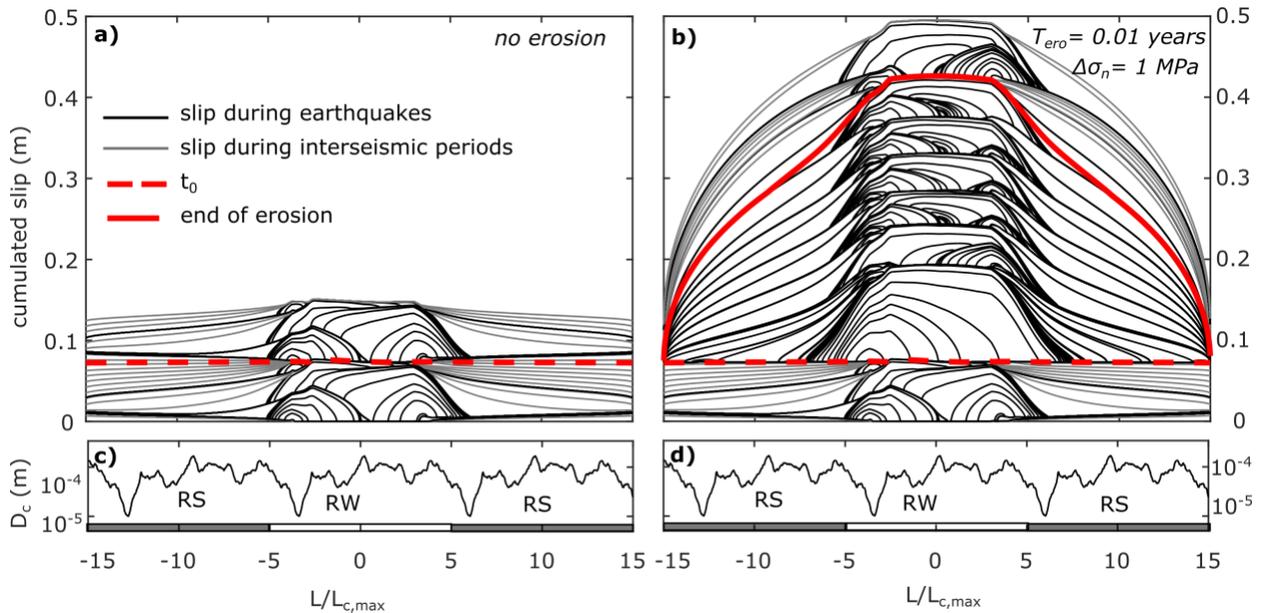

**Figure 2**. Cumulated slip along strike during four years for a) the undisturbed fault and b) the fault under $\Delta\sigma_n = 1$ MPa and $T_{ero} = 0.01$ years. The slip is plotted every 0.5 sec during seismic events and every 0.2 years during interseismic periods. The slip at time $t_b$ is plotted in dashed red line, and the slip at $t_e$ is plotted in plain red line in b). c) and d): $D_c$ distribution along the rate-strengthening (RS) and rate-weakening (RW) areas.

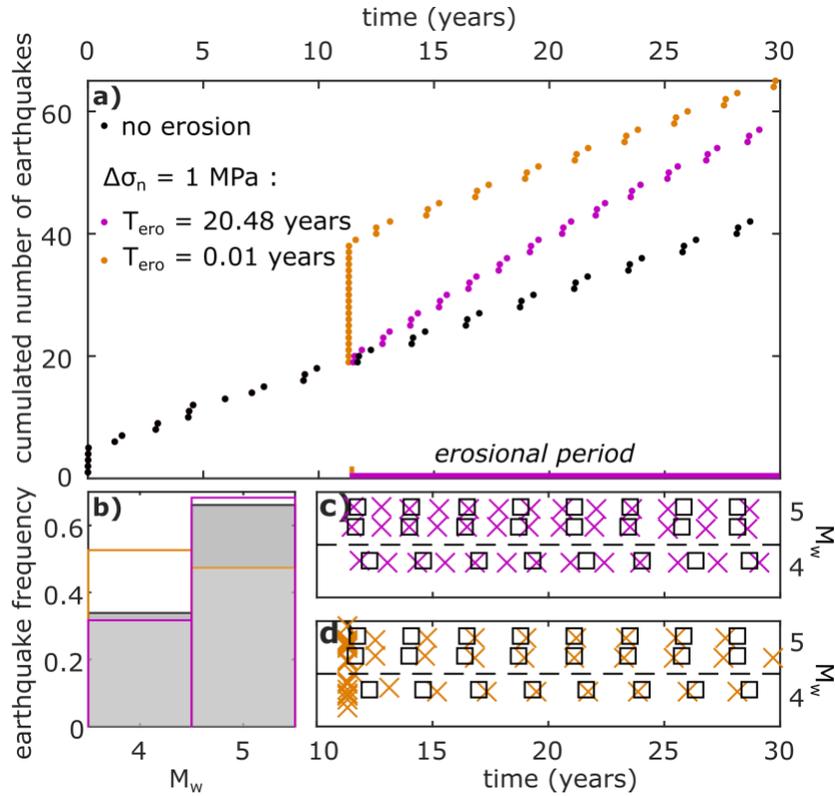

**Figure 3**. a) Cumulated number of earthquakes for $T_{ero} = 0.01$ year and 20.48 years with $\Delta\sigma_n = 1$ MPa, and for the undisturbed fault. b) Proportion of small and large earthquakes during the erosional period (coloured edges), for the two scenarios shown in a). The grey bars show the earthquake distribution for the undisturbed fault. Panels c) and d) show time evolution of earthquakes magnitudes for $T_{ero} = 20.48$ and 0.01 years, compared to the undisturbed fault (dark squares). The horizontal dotted lines show the edge of the bins used in b).

The earthquake frequency also increases with increasing $\Delta\sigma_n$ (Fig. 4c). Compared to the reference case without erosion, it increases by a factor of 1 to 2 for $\Delta\sigma_n = 0.01$ MPa, 1 to 100 for $\Delta\sigma_n = 0.1$ MPa and 1 to 10000 for $\Delta\sigma_n = 1$ Mpa. For a given $\Delta\sigma_n$, earthquake rate increases with decreasing $T_{ero}$. When erosion is shorter than a seismic cycle, earthquake frequency increases significantly, by a factor of 2 for $T_{ero}/T_{cycle} < 2$ with $\Delta\sigma_n = 1$ MPa, or $T_{ero}/T_{cycle} < 0.5$ with $\Delta\sigma_n = 0.1$ or 0.01 MPa.

The proportion of large ($M_w > 4.5$) and small ($M_w < 4.5$) earthquakes during erosion varies as shown for the models with $\Delta\sigma_n = 1$ MPa (Fig. 4d). For $T_{ero} > 10 \, T_{nuc}$, the size distribution of earthquakes does not vary significantly from the distribution of the undisturbed fault (when large earthquakes represent 2/3 of all rupture events). However, for $T_{ero} < 10 \, T_{nuc}$ the proportion of earthquakes is inverted with a significant increase of small events and a decrease of the larger ones. This variation is not observed for small $\Delta\sigma_n$ (0.1 and 0.01 Mpa) (fig. S2).

## 4 Discussion and concluding remarks

Using quasi-dynamic models of earthquake cycle on a mode II fault, we here show that a large erosional event simulated by a variation of the normal stress $\Delta\sigma_n$ over a certain time $T_{ero}$ can significantly affect earthquake statistics.

An erosional event can result in a clear increase in earthquake frequency. For large $\Delta\sigma_n$, the fault response is quite simple as earthquake frequency increases with the rate of normal stress change (i.e. decreasing $T_{ero}$). For smaller $\Delta\sigma_n$, our results illustrate the complexity of fault response to transient stress changes. For example, stress variation with low magnitudes ($\Delta\sigma_n = 0.01$ MPa) occurring within a too short period do not trigger any earthquake (fig. 4a). If the same total normal stress is removed over a longer period, it triggers aseismic, or seismic slip during erosion (fig. 4b). This suggests that within a population of faults close to their critical state, even small normal stress variations could trigger numerous earthquakes within the years following a large erosional event.

In our model (equation 2), normal stress variation itself does not contribute to the evolution of frictional state (e.g., Linker & Dieterich, 1992). Although we do not expect the Linker-Dieterich effect to significantly change our results, we suspect that it would enhance the erosion-induced shear strength decrease, and then the modelled seismicity. Moreover, poroelasticity, thermal pressurization and dilatant strengthening act on the fault strength and could also enhance seismicity or compete with each other (e.g., Segall et al., 2010). Surface unloading would also induce shear stress variations. In the case of a reverse fault, we could expect the erosion-induced shear stress increase (Steer et al., 2014) to enhance the observed earthquake production (e.g., Hawthorne & Rubin, 2013; Johnson et al., 2017; Luo & Liu, 2019).

We also show that under high and rapid enough $\Delta\sigma_n$, a single fault is likely to produce more numerous small ruptures, relative to large ones. This observation could be biased by our chosen set-up. First, the lowest Dc areas are located at the edge of the RW patch (Fig. 1b-c). However, Ader et al., 2014 noted a similar change in the distribution of events following a step-like increase in shear stress on an homogeneous fault. Moreover, running the same model on a fault with another random pattern of $D_c$ with same roughness leads to the same observation (fig. S3, supporting information). The correlation length also probably influences the overall distribution of seismicity. This effect needs to be further investigated.

We thus suspect that under a larger normal stress decrease, the reduction of slip induced by a smaller shear stress drop associated with a reduction in critical stiffness (e.g., Leeman et al., 2016) contributes to reducing the proportion of large earthquakes. Moreover, rapid normal stress variations, along with a spatially heterogeneous $D_c$, could significantly change nucleation length scales allowing for the fault to rupture with multiple smaller ruptures than the canonical case. To confirm our hypothesis, 3D modelling and simulation of a wide range of magnitudes could be carried out (Hillers et al., 2007; Luo et al., 2017).

The earthquake rate increases even for very small $\Delta\sigma_n$ as long as $T_{ero}$ is short enough compared to the duration of an undisturbed seismic cycle. The dependency of fault response to the magnitude and frequency of environmental stress change has already been documented.

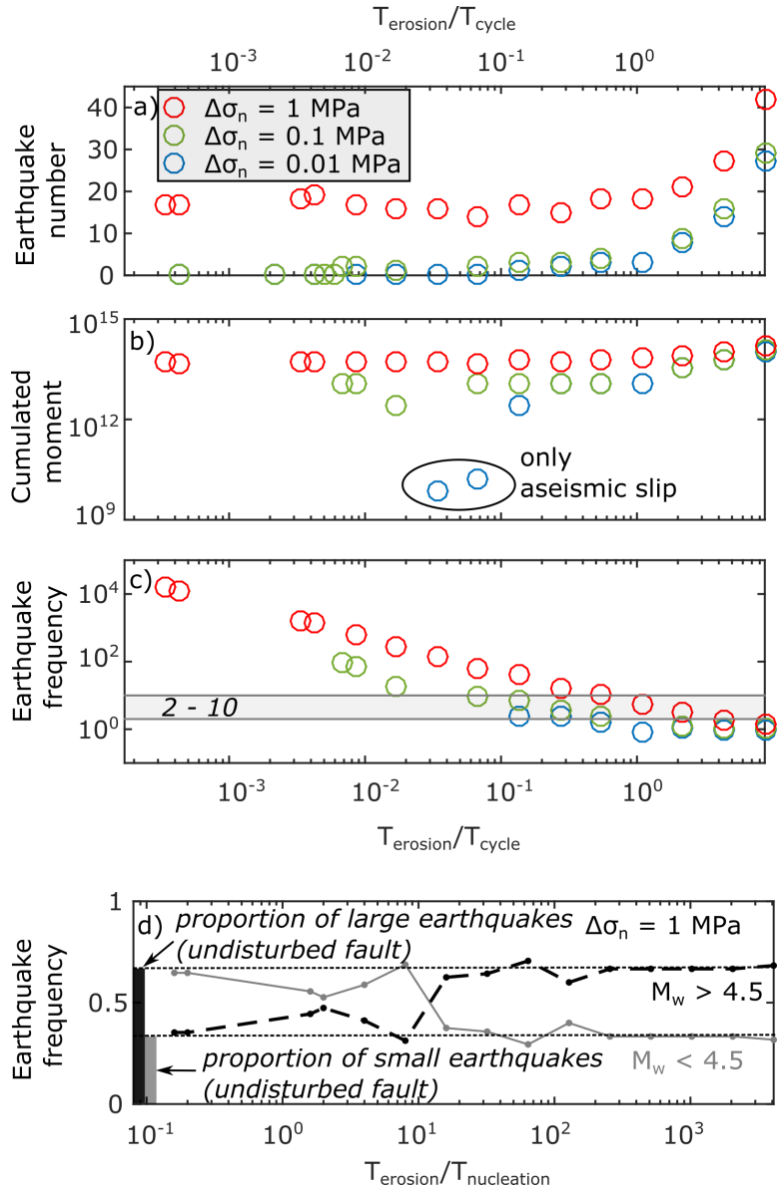

**Figure 4**. a) Number of earthquakes N during the erosional event, as a function of $T_{ero}$ normalised by the duration of an undisturbed seismic cycle. b) Cumulated moment during erosion. The two models quoted 'aseismic slip' correspond to two scenarios in which no earthquake occurred. c) Earthquake rate during the erosional period, normalised by the rate of the undisturbed fault. The shaded area shows the models for which the seismicity rate increases by a factor of 2 to 10 compared to the undisturbed seismicity rate. d) Proportion of large ($M_w > 4.5$, dark line) and small ($M_w < 4.5$, grey line) earthquakes during erosion for each model with $\Delta\sigma_n = 1$ MPa, as a function of $T_{ero}$ normalised by the earthquake nucleation time. The dotted lines show the proportion of small and large earthquakes in the case of the undisturbed fault.

Earthquake-induced sudden stress changes below 0.01 MPa were observed to be insufficient to trigger seismicity (Hardebeck et al., 1998; Reasenberg & Simpson, 1992). Hawthorne & Rubin, (2013) have demonstrated tidal modulation of slow slip events. However, earthquake rate does not systematically display variations at tidal period (Cochran et al., 2004; Vidale et al., 1998), despite the similar magnitude of static stress changes due to hydrological cycle and Earth's tides.

Such period-dependency has also been observed in laboratory experiments (Beeler & Lockner, 2003; Lockner & Beeler, 1999). For high frequencies, the fault response is amplitude-dependent while for low frequencies, it rather depends on perturbation frequency and amplitude (Boettcher, 2004). This transition has been interpreted as the time necessary to reach the critical distance $D_c$. Numerical modelling (Ader et al., 2014) shows a resonance effect in the response of a finite fault to harmonic shear stress variations, which is more important than for 1D spring slider models with rate and state friction laws (Perfettini et al., 2001). Kaneko & Lapusta, (2008) and Ader et al. (2014) pointed out similar observations studying a finite fault response to a static shear stress step.

In this study, we show that the fault response to one transient stress change is also period-dependent. We observe a range of erosion periods for which a normal stress variation of 0.1 to 1% can significantly accelerate seismicity on our modelled fault. This range is bounded by the typical time-scales of the modelled seismic cycle and earthquake nucleation. Fault response is likely to depend on plate velocity, which is inversely proportionally related to the recurrence time of earthquakes (Ader et al., 2014). Analogously, Ader et al. (2014) noted an inversely proportional relationship between the loading rate and the characteristic response time of seismicity, when considering either sinusoidal or step-like stress variation. Therefore, our finding that fault response is greater for an erosion period related to the recurrence time of large earthquake in our model should not change using a different loading rate.

In nature, landscape response to large erosional events is likely to occur at a time-scale ranging from a few years to several decades (Hovius et al., 2011; Howarth et al., 2012; Croissant et al., 2017). Moreover, sediment export is expected to be most efficient and to significantly exceed background erosion rates over the first years following the perturbation (Hovius et al., 2011; Croissant et al., 2017). Moreover, earthquake nucleation takes months to a year (Savage et al., 2007, Beeler et al., 2003), and seismic cycles last between about 100 to 1000 years (e.g., Chen et al., 2007). Hence, the timescale of an erosional event ranges between the nucleation and the seismic cycle timescales. Therefore, our results suggest that one large erosional event is likely to increase seismicity by at least a factor of two, if it implies normal stress decrease of at least 0.1% from the background normal stress. For example, overpressured faults with a normal stress of about 25 MPa below 2 km depth (Suppe 2014) would be sensitive to an erosional event of a few decades up to 5 km depth, considering the induced static stress change (Steer et al., 2014). This corroborate previous observation of an increase in earthquake frequency by a factor of two and a b-value increase in the years following typhoon Morakot (Steer et al., 2020).

Normal stress change due to erosion is different from a sudden static shear stress change induced by a mainshock, because it is likely to be transient. However, contrary to hydrological, tidal or atmospheric forcing, surface processes such as erosion and sedimentation are not periodic.

Therefore, the induced stress changes are likely to cumulate over time. By showing that erosion can significantly trigger seismicity at seismic cycle time-scale, our results build upon previous results showing the impact of erosion on static stress changes (Steer et al., 2014). They also suggest that such cumulative processes, including large erosional events, but also glacial melting, or human-induced water extraction, can significantly contribute to the deformation of the crust at least in its shallow part.

## Acknowledgments

This research has been supported by the Agence Nationale de la Recherche (grant no. ANR-14-CE33-0005). We thank two anonymous reviewers for their constructive comments. We also thank Dimitri Lague, Philippe Davy and Pierre Romanet for fruitful discussions. The simulations were performed using QDYN v1.1 (Luo et al., 2017), publicly available at https://github.com/ydluo/qdyn.

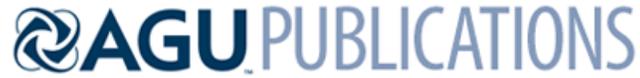

**The Impact of Large Erosional Events and Transient Normal Stress Changes on the Seismicity of Faults**

L. Jeandet Ribes[1,2], N. Cubas[1], H. S. Bhat[3], and P. Steer[2]

[1] Institut des Sciences de la Terre Paris, ISTeP UMR 7193, Sorbonne Université, CNRS-INSU, 75005 Paris, France

[2] Univ Rennes, CNRS, Géosciences Rennes - UMR 6118, 35000 Rennes, France

[3] Laboratoire de Géologie, Ecole Normale Supérieure, CNRS-UMR 8538, PSL Research University, Paris, France

**Contents of this file**

**Introduction**

This supplementary file gives more details about the numerical method used to model the effects of a normal stress decrease on seismicity (texts **S1** and **S2**) and to compute the earthquake catalogue (text **S3** and figure **S1**). Figure **S2** shows the proportion of small and large earthquakes for the cases $\Delta\sigma_n = 10^5$ Pa and $\Delta\sigma_n = 10^4$ Pa. Figure **S3** shows the results using different Dc pattern. Figure **S4** provides an explanation of the method we use to compute the nucleation time.

**S1. Fault discretization and choice of Dc range**

The size of the cohesive zone $L_b$ (Rice et al., 1983; Lapusta et al., 2009) is defined as:

$$L_b = \frac{GD_c}{b\sigma} \qquad (S1)$$

and the characteristic length for nucleation (Rubin et al., 2005) as:

$$L_c = \frac{2bGD_C}{\sigma\pi(b-a)^2} \qquad (S2)$$

In our model setup, $D_c$ varies along strike. The seismogenic patch needs to be larger than the largest $L_c$ value $L_{c,max}$ in order to allow the full propagation of dynamic ruptures. The cell size must be smaller than the smallest $L_b$ value $L_{b,min}$ in order to correctly capture the model response. Therefore, the number of cells in the model depends directly on the ratio between maximum and minimum $D_c$. Then, the range of $D_c$ must satisfy a compromise between the range of modelled earthquake magnitudes and the computational time.

We chose to set the length of the seismogenic patch to 10 $L_{c,max}$ in order to get dynamic events, and we ensure at least 8 cells per minimum $L_b$ unit. Then, we vary $D_C$ between $D_{c,max} = 3.4 \times 10^{-4}$ m ($L_{c,max}$ = 561 m) and $D_{c,min} = 2 \times 10^{-5}$ m ($L_{c,min}$ = 4.2 m). This ratio of 17 between maximum and minimum $D_C$ values leads to a discretization into 32768 cells. It allows us to model seismic moments covering one order of magnitude, within a reasonable computational time (one week for a typical simulation of 30 years).

## S2. Implementation of normal stress change

In QDYN, the time derivative of the equilibrium equation is solved for velocity, for a constant normal stress:

$$\frac{d\tau}{dt} - \zeta \frac{dV}{dt} = \sigma_n \left( \frac{\partial \mu}{\partial V} \frac{\partial V}{\partial t} + \frac{\partial \mu}{\partial \theta} \frac{\partial \theta}{\partial t} \right) \qquad (S3)$$

Where $\tau$ is the elastic shear stress due to slip, $V$ is the velocity on the fault, $\zeta$ is the fault impedance (radiation damping term), $\sigma_n$ is the normal stress on the fault, $\mu$ is the friction coefficient, and $\theta$ the state variable of the friction rate-and-state law.

We modified it to include a time-dependant normal stress:

$$\frac{d\tau}{dt} - \zeta \frac{dV}{dt} = \sigma_n(t) \left( \frac{\partial \mu}{\partial V} \frac{\partial V}{\partial t} + \frac{\partial \mu}{\partial \theta} \frac{\partial \theta}{\partial t} \right) + \frac{d\sigma_n}{dt} \mu(t) \qquad (S4)$$

The time-dependant normal stress and its derivative are defined as:

- if $t < t_b$ or $t \geq t_e$ : $\sigma_n(t) = \sigma_{n,0}$ and $\frac{d\sigma_n}{dt} = 0$
- if $t_b \leq t < t_e$ :

$$\frac{d\sigma_n}{dt} = 2\Delta\sigma_n \frac{t - t_e}{(t_e - t_b)^2} \qquad (S5)$$

$$\sigma_n(t) = \sigma_{n,0} + \frac{\Delta\sigma_n}{(t_e - t_b)^2} (t_b - t)(2t_e - t_b - t) \qquad (S6)$$

With $\Delta\sigma_n$ the total removed normal stress during erosion ($\Delta\sigma_n > 0$), $t_b$ and $t_e$ the times of beginning and end of erosion, respectively.

## S3. Construction of the earthquake catalogue

To construct the earthquake catalog, we integrate the velocities over the rate-weakening (RW) patch to obtain a linear moment rate:

$$\dot{M}_0(t) = G * Z * \int_{RW} V(x,t)\,dx \qquad (S7)$$

where V(x) is the velocity at the location x of the fault at time t, G the Young's modulus and Z = 10 km the fault width. We then isolate seismic events when $\dot{M}_0 > 10^8$ dyn.cm$^{-2}$.s$^{-1}$, corresponding to the onset of seismic slip in a typical simulation. Changing this threshold, for example to $10^9$ dyn.cm$^{-2}$.s$^{-1}$, will slightly change the magnitude of individual earthquakes but neither the earthquake frequency nor the distribution of magnitudes (Fig **S1**).

The moment rate is integrated over the full earthquake duration to compute a classical moment magnitude:

$$M_w = \frac{2}{3} * \log(M_0) - 6.07 \qquad (S8)$$

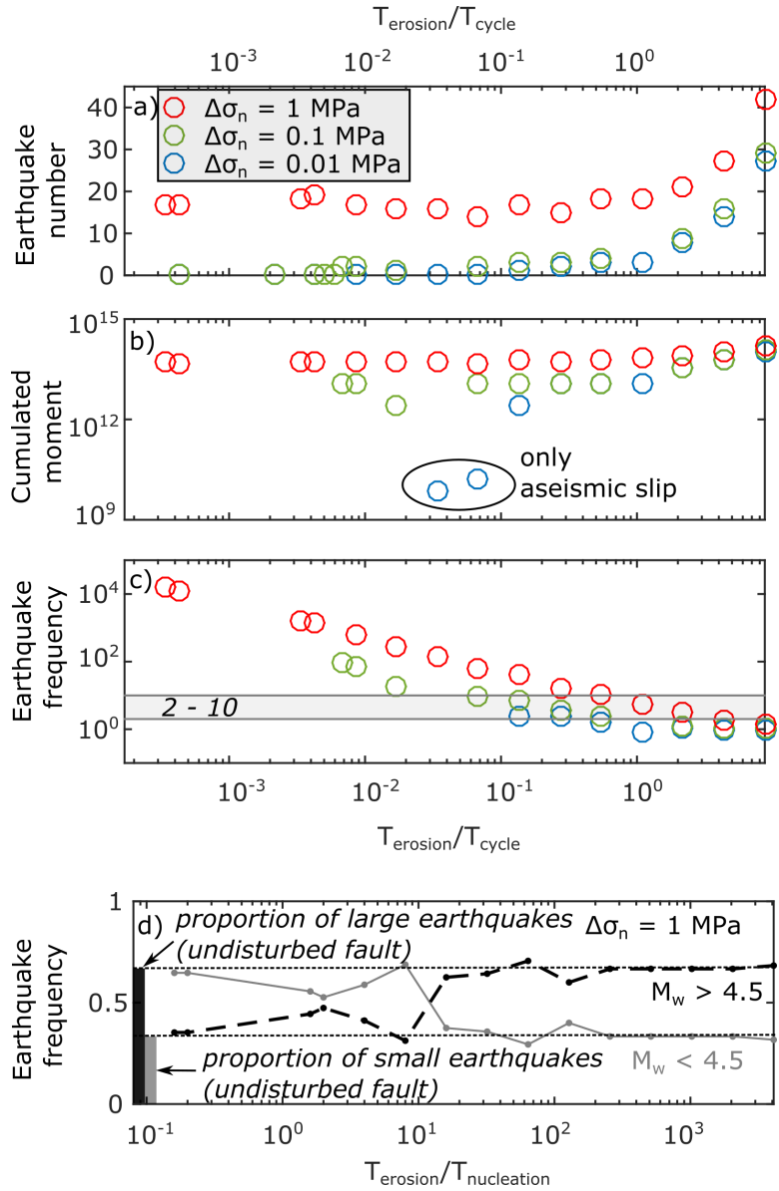

**Figure S1.** – Same as figure 4 in the main text, but using a moment rate threshold $\dot{M}_0 > 10^9$ dyn.cm$^{-2}$.s$^{-1}$ instead of $\dot{M}_0 > 10^8$ dyn.cm$^{-2}$.s$^{-1}$.

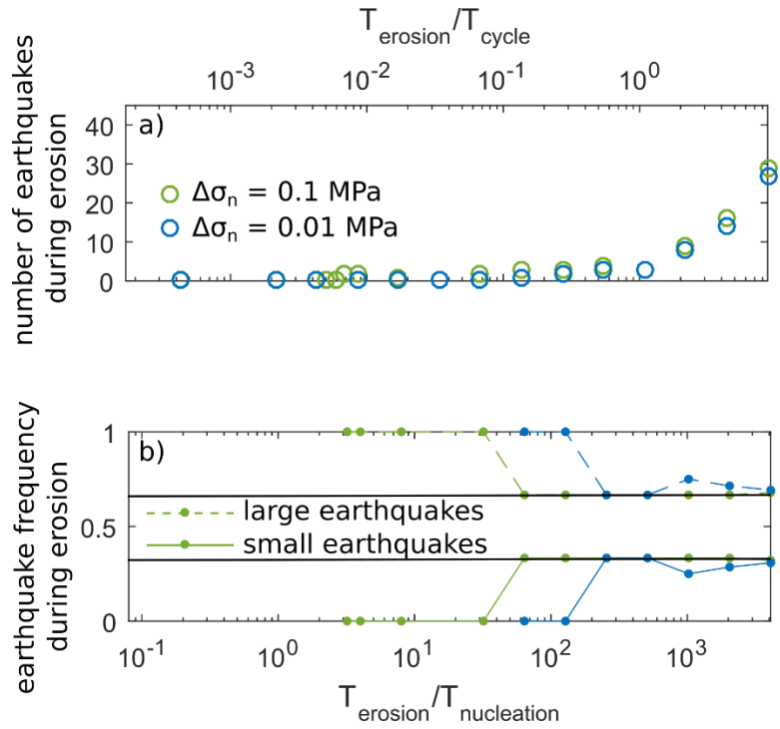

**Figure S2**. – Same as figure 4a) and 4d) of the main text, with $\Delta\sigma_n = 10^5$ Pa and $\Delta\sigma_n = 10^4$ Pa. In b), the dark lines show the proportion of large and small earthquakes in the case of $\Delta\sigma_n$ = zero.

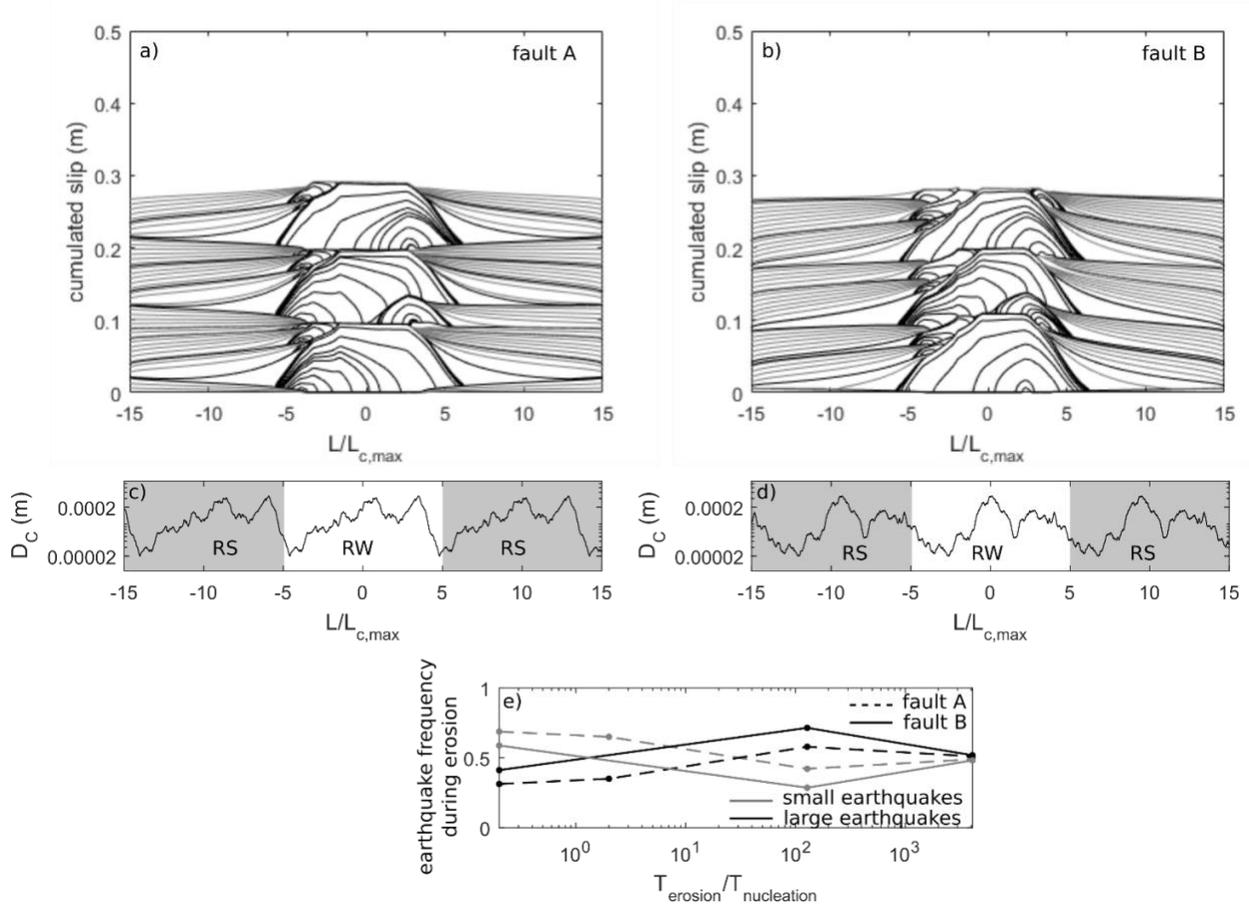

**Figure S3.** – Results using a $D_c$ pattern different from the fault modelled in the main paper. a) and b) show the cumulated slip during 8.5 years in the case of the undisturbed fault. Aseismic and seismic slip are plotted using the same time intervals than in the main text. Different earthquake cycles occur resulting from the different $D_c$ distribution (c and d). e) Proportion of small ($M_w < 4.5$) and large ($M_w > 4.5$) earthquakes during the erosional period ($\Delta\sigma_n = 10^6$ Pa) as a function of $T_{ero}$ normalized by the nucleation time of a typical earthquake (4 models for fault A, 3 models for fault B). In both scenarios, the proportion of small earthquakes during erosion increases for small erosional periods.

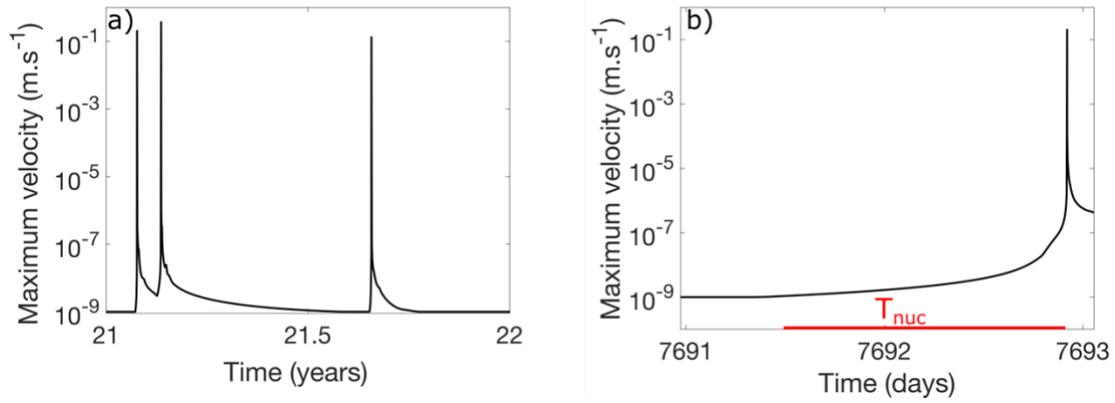

*Figure S4 - Time evolution of fault maximum velocity for a sequence of three characteristic earthquakes (a) and zoom on the initiation of slip for the first earthquake of the sequence (b). We compute the nucleation time as the time between the onset of slip acceleration and the reaching of seismic velocity.*